\documentclass[preprintnumbers, floatfix, preprintnumbers, letterpaper, superscriptaddress,nofootinbib, twocolumn]{revtex4}
\pdfoutput=1
\usepackage{graphicx}
\usepackage{microtype}
\usepackage{amsmath}
\usepackage{amssymb}
\usepackage{subfigure}
\usepackage{hyperref}
\usepackage{url}
\usepackage{xcolor}
\usepackage{color}
\usepackage{mathrsfs}
\usepackage{calrsfs}
\usepackage{amsfonts}
\usepackage{latexsym}
\usepackage{ragged2e}
\usepackage{epsfig}
\usepackage{textcomp}
\usepackage{phaistos}
\makeatletter
\renewcommand\@makefnmark{\hbox{\@textsuperscript{\normalfont\color{purple}\@thefnmark}}}
\renewcommand\@makefntext[1]{%
  \parindent 1em\noindent
            \hb@xt@1.8em{%
                \hss\@textsuperscript{\normalfont\@thefnmark}}#1}
\makeatother

\usepackage{caption}
\DeclareCaptionJustification{justified}{\leftskip=0pt \rightskip=0pt \parfillskip=0pt plus 1fil}
\definecolor{vividviolet}{rgb}{0.62, 0.0, 1.0}
\definecolor{amaranth}{rgb}{0.9, 0.17, 0.31}
\definecolor{palatinateblue}{rgb}{0.15, 0.23, 0.89}
\definecolor{brightpink}{rgb}{1.0, 0.0, 0.5}
\definecolor{cornflowerblue}{rgb}{0.39, 0.58, 0.93}
\definecolor{deepcarminepink}{rgb}{0.94, 0.19, 0.22}
\definecolor{radicalred}{rgb}{1.0, 0.21, 0.37}

\hypersetup{ linktoc=all,
    colorlinks, linkcolor={palatinateblue},
    citecolor={brightpink}, urlcolor={amaranth}
}

\graphicspath{{Images/}}

\renewcommand{\d}[1]{\ensuremath{\operatorname{d}\!{#1}}}

\graphicspath{{Images/}}
\renewcommand{\d}[1]{\ensuremath{\operatorname{d}\!{#1}}}
\makeatletter
\def\@fnsymbol#1{\ensuremath{\ifcase#1\or $\textleaf$ \or $\PHplaneTree$
\else\@ctrerr\fi}}%

\makeatother

\def\sideremark#1{\ifvmode\leavevmode\fi\vadjust{\vbox to0pt{\vss
 \hbox to 0pt{\hskip\hsize\hskip1em
 \vbox{\hsize1.5cm\tiny\raggedright\pretolerance10000
 \noindent #1\hfill}\hss}\vbox to8pt{\vfil}\vss}}}%
\begin{document}

\title{The Attractor of Evaporating Reissner-Nordstr\"om Black Holes}

\author{Yen Chin \surname{Ong}}
\email{ycong@yzu.edu.cn}
\affiliation{Center for Gravitation and Cosmology, College of Physical Science and Technology, Yangzhou University, \\180 Siwangting Road, Yangzhou City, Jiangsu Province  225002, China}
\affiliation{School of Aeronautics and Astronautics, Shanghai Jiao Tong University, Shanghai 200240, China}

\begin{abstract}

Hiscock and Weems showed that there is an attractor behavior in the evolution of asymptotically flat Reissner-Nordstr\"om black hole under Hawking evaporation. If the initial charge-to-mass ratio $Q/M$ of the black hole is relatively small, then the ratio first increases until the black hole hits the attractor, and then starts to discharge towards the Schwarzschild limit. Sufficiently charged black holes, on the other hand, simply discharge steadily towards the Schwarzschild limit. In this work we further investigate the nature of the attractor, and found that it is characterized by the mass loss rate being equal to the charge loss rate.
The attractor is not necessarily related to the specific heat in a general evaporating black hole spacetime, but for the Reissner-Nordstr\"om case part of the attractor lies very close to the boundary of the region where specific heat changes sign.
\end{abstract}

\maketitle

\section{Introduction: Attractor in the Charged Black Hole Evolution}\label{1}

Black holes can lose its mass, angular momentum, and charge by emitting Hawking radiation. However,
For an asymptotically flat Schwarzschild black hole, the usual model that assumes thermality holds implies that the evolution of the mass follows the Stefan-Boltzmann law:
\begin{equation}\label{SB}
\frac{\d M}{\d t}= -\alpha a \sigma T^4,
\end{equation}
where we have set the Boltzmann constant $k_B=1$. Here $a=\pi^2/(15\hbar^3)$ is the radiation constant, $T$ the Hawking temperature, and $\sigma=27\pi M^2$ is the effective area whose radius corresponds to the impact parameter of the photon orbit at $r=3M$ in the geometric optics approximation. Due to scattering at long wavelengths, the effective emission surface is actually smaller. This is governed by $\alpha$, the greybody factor, which depends on the species of the emitted particles. Unless one is concerned with the exact lifetime, the factor $\alpha$ can be neglected. We are only interested in the general feature of black hole evolution, for this reason we will set $\alpha=1$ in this work. From this simple differential equation we can obtain the standard result that a Schwarzschild black hole takes a finite time proportional to the cube if its initial mass to completely evaporate. 

On the other hand, even inclusion of just one other parameter (namely the electrical charge or the angular momentum), can render the evolution rather nontrivial. For example, evaporating black holes can spin \emph{up}, and the end state of Hawking evaporation need not be Schwarzschild. Instead, the fate of a rotating black hole depends on the number and species of particles being emitted \cite{9710013, 9801044}. See also the appendix of \cite{1210.6348}. While it is not so surprisingly that rotating black hole spacetime can behave in a complicated manner, what \emph{is} surprising is that even \emph{non-rotating} charged black holes can have nontrivial behavior under Hawking evolution.
In \cite{HW}, Hiscock and Weems showed that for sufficiently large Reissner-Nordstr\"om black hole, the charge-to-mass ratio is not monotonic in time. 

Specifically, the metric of a Reissner-Nordstr\"om black hole is, in the units such that $G=c=4\pi \epsilon_0=1$,
\begin{flalign}
g[\text{RN}]=&-\left(1-\frac{2M}{r}+\frac{Q^2}{r^2}\right)\d t^2\\ \notag & + \left(1-\frac{2M}{r}+\frac{Q^2}{r^2}\right)^{-1}\d r^2 + r^2\d\Omega^2_{S^2}.
\end{flalign}
Its event horizon is located at $r_+=M+\sqrt{M^2-Q^2}$ and its Hawking temperature is\footnote{We follow the convention of \cite{HW} in which $\hbar \neq 1$, but rather it is the Planck area. Consequently temperature has dimension of length.}
\begin{equation}\label{temp}
T=\frac{\hbar\sqrt{M^2-Q^2}}{2\pi \left(M+\sqrt{M^2-Q^2}\right)^2}.
\end{equation}
A larger black hole has a lower temperature, and thus is ``less energetic'' to pair-produce massive particles in its Hawking emission. Therefore for sufficiently large black holes, it suffices to consider massless particles and the lightest charged particle, namely the electron/positron. In this regime, charge loss can be approximated by the Schwinger formula \cite{HW, g}:
\begin{equation}\label{dQdt}
\frac{\d Q}{\d t} \approx -\frac{e^4}{2\pi^3\hbar m^2}\frac{Q^3}{r_+^3}\exp\left(-\frac{r_+^2}{Q_0Q}\right),
\end{equation}
where $Q_0$ is the inverse of the Schwinger critical field: $E_c:={m^2 c^3}/{e\hbar} = 1.312 \times 10^{16}~\text{V/cm}$. 
Here $m$ and $e$ denote the mass and the charge of the positron, respectively.
The Hiscock-Weems model works for sufficiently large black holes $M \gg Q_o := \hbar e/(\pi m^2)$ \cite{HW}. 

Mass loss is due to the emission of neutral massless particles following the Stefan-Boltzmann law, Eq.(\ref{SB}), and the emission of electron/positron via $\d Q/\d t$ term, which enters via the first law of black hole thermodynamics:
\begin{equation}\label{dMdt}
\frac{\d M}{\d t} = -a \alpha \sigma T^4 + \frac{Q}{r_+}\frac{\d Q}{\d t}.
\end{equation}
Here, the effective emission area, which can be calculated from the metric is \cite{HW} 
\begin{equation}
\sigma[\text{RN}]=\frac{\pi}{8}\frac{(3M+\sqrt{9M^2-8Q^2})^4}{(3M^2-2Q^2+M\sqrt{9M^2-8Q^2})}.
\end{equation}
Assuming without loss of generality that $Q>0$ (and so we are dealing predominantly with positrons),
 we can study how the charge-to-mass ratio $Q/M$ evolves by solving this system of coupled ordinary differential equations numerically.

\begin{figure}[!h]
\centering
\includegraphics[width=3.4in]{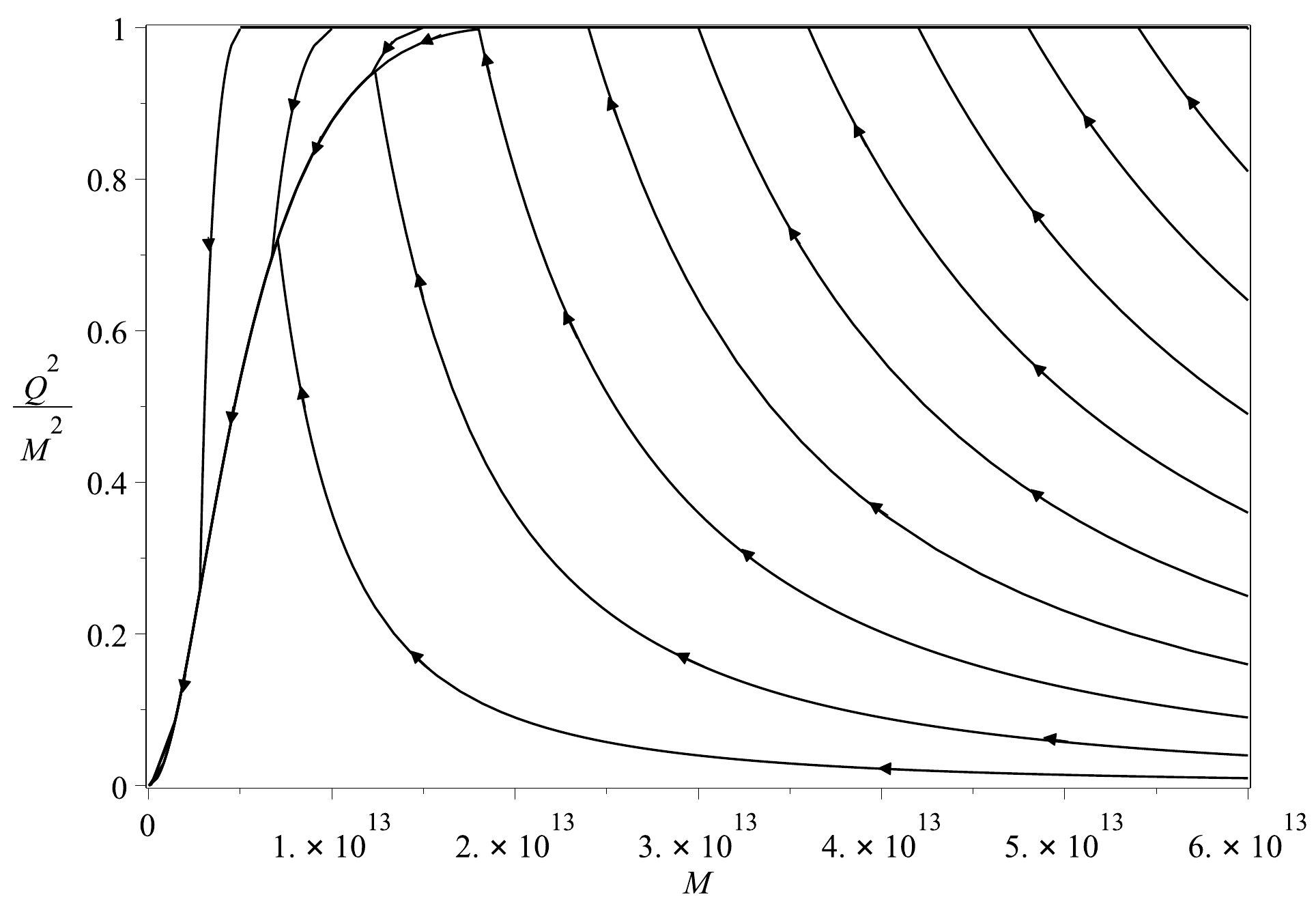}
\caption{The plot of $(Q/M)^2$ against $M$ of evaporating Reissner-Nordstr\"om black holes, where both $M$ and $Q$ are both functions of time governed by Eq.(\ref{dQdt}) and Eq.(\ref{dMdt}). That is, $M$ can be treated as a proxy for $t$: increasing $t$ is decreasing $M$. The arrows denote the direction of time evolution. The mass is measured in centimeters.
\label{HWfig}}
\end{figure}

In Fig.(\ref{HWfig}) we plotted $(Q/M)^2$ against $M$, 
and we observe that there are a few interesting features. Firstly, for black holes with a relatively small amount of charge, the ratio $Q/M$ would first increase. For some initial conditions this ratio can approach unity (extremal value), but it never quite reaches it (thus satisfying the third law of black hole thermodynamics, as well as the cosmic censorship). In fact, we see that there is an attractor behavior: once the curve hits the attractor it will flow along it\footnote{We used the word ``hit'' colloquially here. The curves can come very close together near the attractor, but none of the curves ever intersect. After all, solutions of ODE is unique.} -- downward -- toward the Schwarzschild limit. Highly charged black holes, on the other hand, simply discharges steadily towards the Schwarzschild limit.
A natural question to ask is: \emph{what characterizes the attractor curve?} 

\section{The Specific Heat and the Attractor}

In \cite{HW}, Hiscock and Weems argued that the attractor arises from the change of sign in the specific heat of the evaporating black hole. 

It is well-known that for an asymptotically flat Schwarzschild black hole, its specific heat is always negative. For a \emph{static} Reissner-Nordstr\"om black hole, it was shown by Davies \cite{davies} that the specific heat is negative if $(Q/M)^2<3/4$ but becomes positive if $3/4 < (Q/M)^2 <1$. For an evaporating black hole, the specific heat $C := \d M/\d T$ can be calculated via  $\d M/\d T=(\d M/\d t)(\d T/\d t)^{-1}$. Since $\d M/ \d t < 0$, the sign of the specific heat is given by the opposite sign of $\d T/\d t$, whose explicit expression is given by\footnote{
Here one employs the chain rule
\begin{equation}
\frac{\d T}{\d t}=\frac{\partial T}{\partial M}\frac{\d M}{\d t} + \frac{\partial T}{\partial Q}\frac{\d Q}{\d t}.
\end{equation}
The explicit expression can  then be obtained from Eq.(\ref{temp}), Eq.(\ref{dQdt}), and Eq.(\ref{dMdt}). 
Note that \cite{HW} contains a typo in the second term in the denominator: the square is missing in $3840 \pi^2$.} 

\begin{equation}
\frac{\d T}{\d t}=\frac{e^4}{4 \pi^4 m^2} \frac{Q^4}{r_+^6} \exp\left(\frac{-r_+^2}{QQ_0}\right)
 - \frac{\hbar^2\alpha}{3840 \pi^2}{f}(M,Q),
\end{equation}
where $f(M,Q)$ is given by the complicated expression
\begin{equation}
\frac{(M^2-Q^2)^{3/2}[3M+(9M^2-8Q^2)^{1/2}]^4[M-2(M^2-Q^2)^{1/2}]}{r_+^{10}[3M^2-2Q^2+M(9M^2-8Q^2)^{1/2}]}.
\end{equation}

\begin{figure}[!h]
\centering
\includegraphics[width=3.4in]{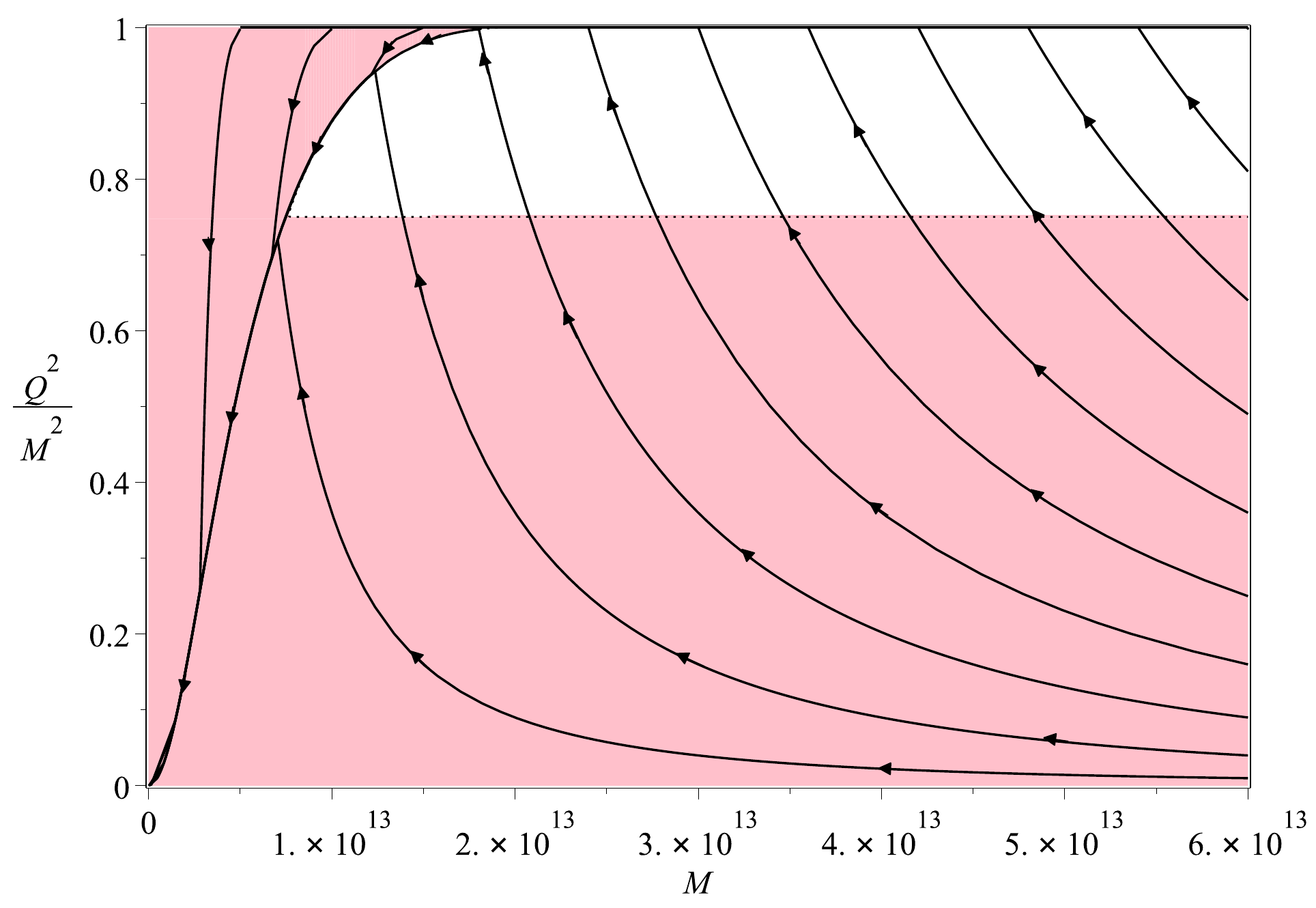}
\caption{Fig.(\ref{HWfig}) with region of negative specific heat indicated in pink shade, while positive specific heat region is in unshaded white. Note that part of the attractor (in the range $3/4<(Q/M)^2<1$) seems to coincide with the boundary of the positive specific heat region.
\label{specific-heat}}
\end{figure}
In Fig.(\ref{specific-heat}), the shaded pink region corresponds to region of negative specific heat, while the unshaded white region at the top right corner is the positive specific heat region. 
As noted by Hiscock and Weems \cite{HW}, the positive specific heat region approaches the result found by Davies \cite{davies} for large enough mass\footnote{Despite the change of sign of specific heat, there is no change to the geometry, and hence there is arguably no real ``phase transition''. See, e.g., \cite{pavon, meitei}.}.

We see that part of the attractor (in the range $3/4<(Q/M)^2<1$) seems to coincide with the boundary of the positive specific heat region. However, this does not explain why the attractor persists down to below $(Q/M)^2<3/4$. Furthermore, this cannot be the full story since the curves that enter the positive specific heat region from below do not change their behavior until they hit the upper boundary of the region (which for large $M$ is close to extremality).  

We can therefore deduce that the positive specific region does \emph{not} imply or characterize the attractor curve.

\section{Mass and Charge Dissipations}

\begin{figure}[!h]
\centering
\includegraphics[width=3.4in]{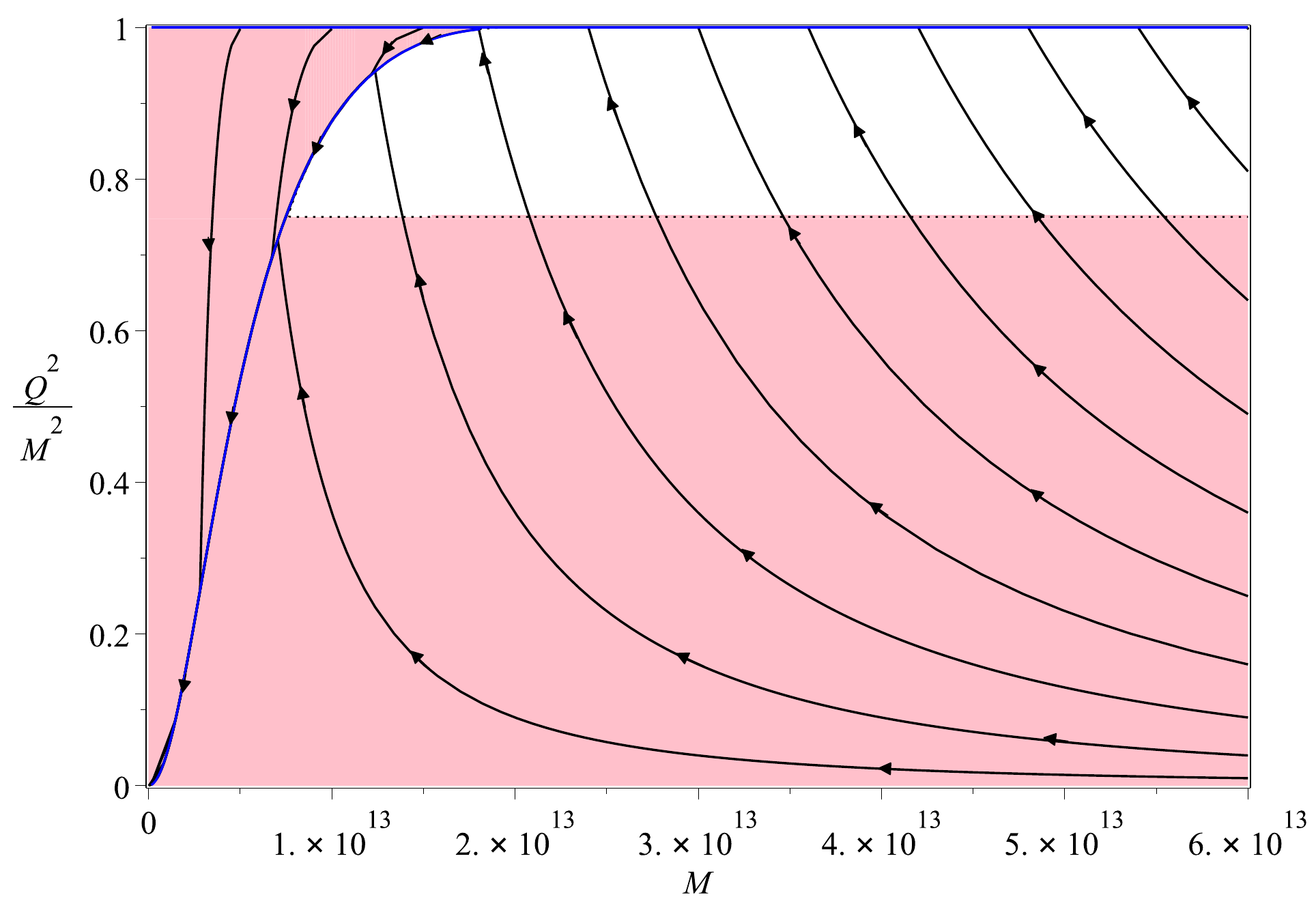}
\caption{The attractor (blue curves) as characterized by the condition $\d M/\d t = \d Q/\d t$, superimposed onto Fig.(\ref{specific-hea}). For small enough $M$, the blue line $Q^2/M^2=1$ becomes a repeller.
\label{realattractor}}
\end{figure}

In fact, as we mentioned, there are two distinct behaviors of charged black holes depending on its initial mass and charge. If the initial charge-to-mass ratio is relatively small, then $Q/M$ increases first until it hits the attractor, because charge loss is relatively inefficient. The black holes are said to be in the mass dissipation regime. On the other hand, if the initial charge-to-mass ratio is relatively large, then $Q/M$ would decrease steadily towards the attractor. These black holes are in the charge dissipation regime. 
Whether $Q/M$ first increases or not depends on whether charge loss or mass loss is more efficient. Thus, it stands to reason that the attractor is when both rate are equal, i.e.,
\begin{equation}\label{cond}
\frac{\d M}{\d t} = \frac{\d Q}{\d t}.
\end{equation}
From Eq.(\ref{dQdt}) and Eq.(\ref{dMdt}), we can obtain the condition 
\begin{flalign}\label{attractorcond}
&\frac{e^4}{2 \pi^3 \hbar m^2}\left(1-\frac{\sqrt{y}}{1+\sqrt{1-y}}\right)\left(\frac{\sqrt{y}}{1+\sqrt{1-y}}\right)^3\\ \notag&\cdot\exp\left[-\frac{M(1+\sqrt{1-y})^2}{Q_0\sqrt{y}}\right]=\frac{a\pi}{8}\frac{M^2(3+\sqrt{9-8y})^4}{3-2y+\sqrt{9-8y}}\\ \notag \cdot &\left(\frac{\hbar \sqrt{1-y}}{2\pi M(1+\sqrt{1-y})^2}\right)^4,
\end{flalign}
which characterizes the attractor. In fact, it was mentioned in \cite{HW} that the neighborhood around the
attractor ``can be roughly described as the region in which the mass-loss rate and the charge-loss rate are of the same order of magnitude''. 
An implicit plot of the condition Eq.(\ref{attractorcond}) indeed yields the attractor curve, which actually has two branches -- one of which is simply the line $y=1$ (which eventually  becomes a ``repeller'' that repels nearby curves away from it when $M$ is small enough). See Fig.(\ref{realattractor}).

\begin{figure}[!h]
\centering
\includegraphics[width=3.4in]{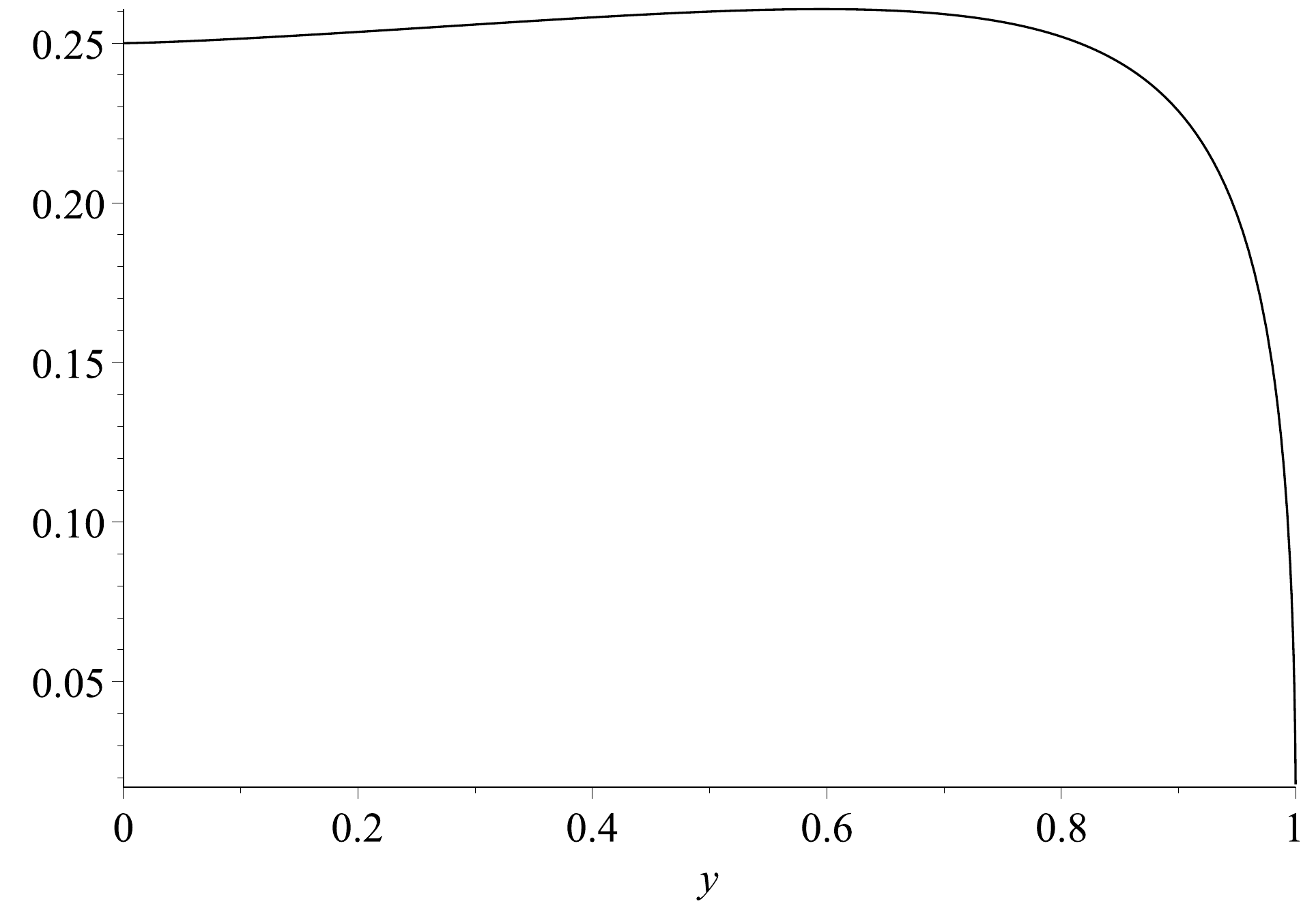}
\caption{The plot of the function $F(y):=\frac{1+\sqrt{1-y}+\sqrt{y}-\sqrt{y}\sqrt{1-y}-2y}{\sqrt{1-y}(1+\sqrt{1-y})^3}$, whose sign is opposite to that of the specific heat.
\label{y}}
\end{figure}

In general, the condition Eq.(\ref{cond}) need not have anything to do with the specific heat. On the attractor curve of the Reissner-Nordstr\"om black hole, however, we have
\begin{flalign}
\frac{\d T}{\d t}&=\frac{\partial T}{\partial M}\frac{\d M}{\d t} + \frac{\partial T}{\partial Q}\frac{\d Q}{\d t} \\ \notag
&=\frac{\d Q}{\d t}\left(\frac{\partial T}{\partial M}+ \frac{\partial T}{\partial Q}\right) \\ \notag
&=-\frac{\hbar}{2\pi M^2}\frac{\d Q}{\d t}\left[\frac{1+\sqrt{1-y}+\sqrt{y}-\sqrt{y}\sqrt{1-y}-2y}{\sqrt{1-y}(1+\sqrt{1-y})^3}\right].
\end{flalign}
Since $\d Q/\d t <0$, the sign of $\d T/\d t$ (which is opposite of that of the specific heat) is therefore the same as the sign of the expression in the square bracket in the last line.
Elementary calculus shows that the function
\begin{equation}
F(y):=\frac{1+\sqrt{1-y}+\sqrt{y}-\sqrt{y}\sqrt{1-y}-2y}{\sqrt{1-y}(1+\sqrt{1-y})^3}
\end{equation}
is indeed positive on the interval $[0,1)$, with a global maximum of $F_\text{max} \approx 0.26065$ at $y \approx 0.59564$. See Fig.(\ref{y}).
This means that the specific heat is still \emph{negative} on the attractor.

In fact, we have
\begin{equation}
\frac{\partial T}{\partial Q} = -\frac{\hbar}{2\pi}\frac{Q(M-\sqrt{M^2-Q^2})}{\sqrt{M^2-Q^2}(M+\sqrt{M^2-Q^2})^3} < 0,
\end{equation}
On the other hand $\partial T/\partial M$ is positive for $y>3/4$ and negative for $y<3/4$, by the result of Davies \cite{davies}. Thus $F(y)$ is trivially positive for $y<3/4$.
As the charge increases to $y>3/4$, $\partial T/\partial M$ changes sign and the function $F$ approaches zero closer and closer. Finally $F(y\to 1)=0$ and the attractor coincides with the boundary where specific heat changes sign.

\begin{figure}[!h]
\centering
\includegraphics[width=3.4in]{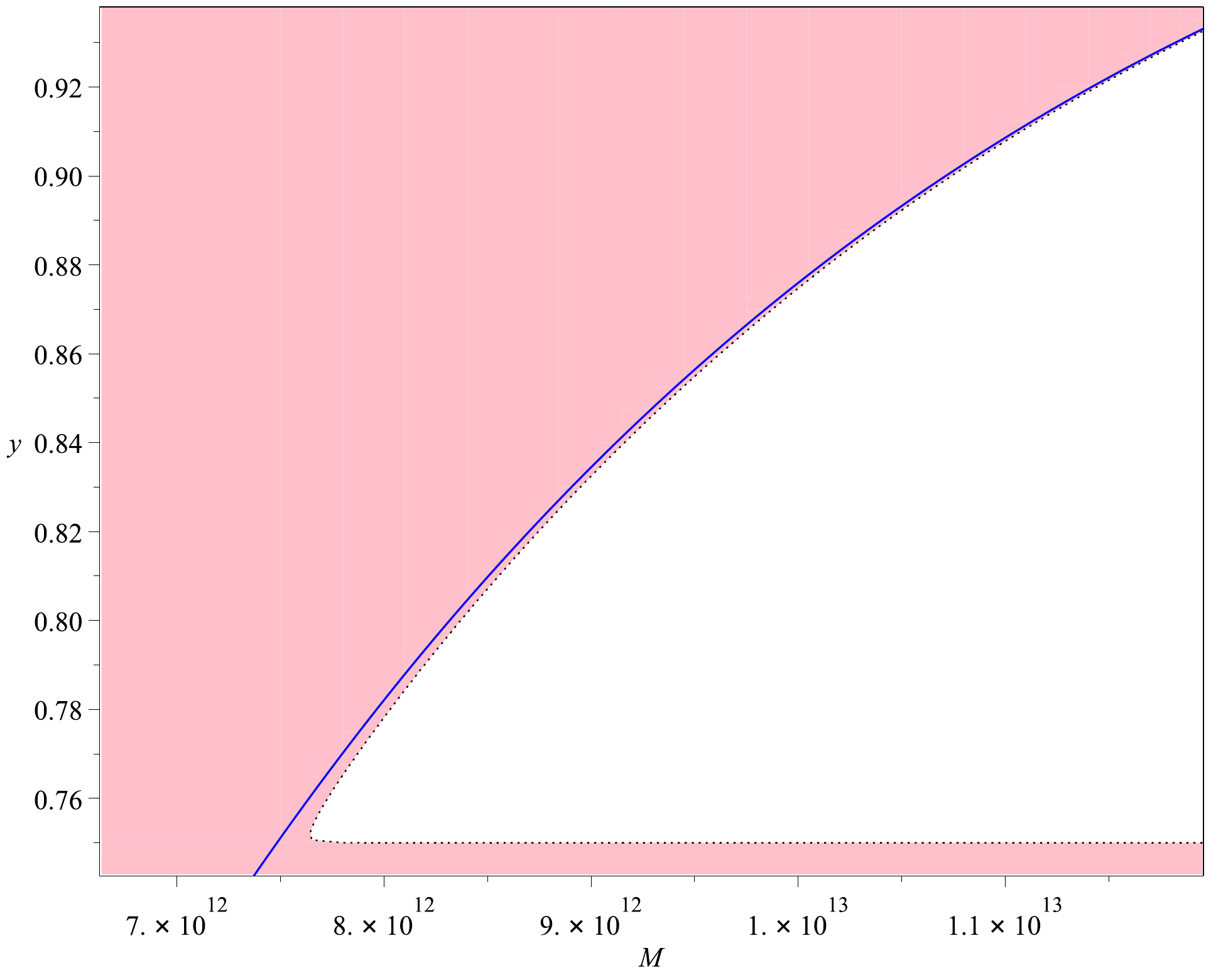}
\caption{The attractor and the boundary of specific heat is quite close but are not equal. The approximation improves near extremality as $y \to 1$, where $y:=(Q/M)^2$.
\label{zoom}}
\end{figure}

\begin{figure}[!h]
\centering
\includegraphics[width=3.4in]{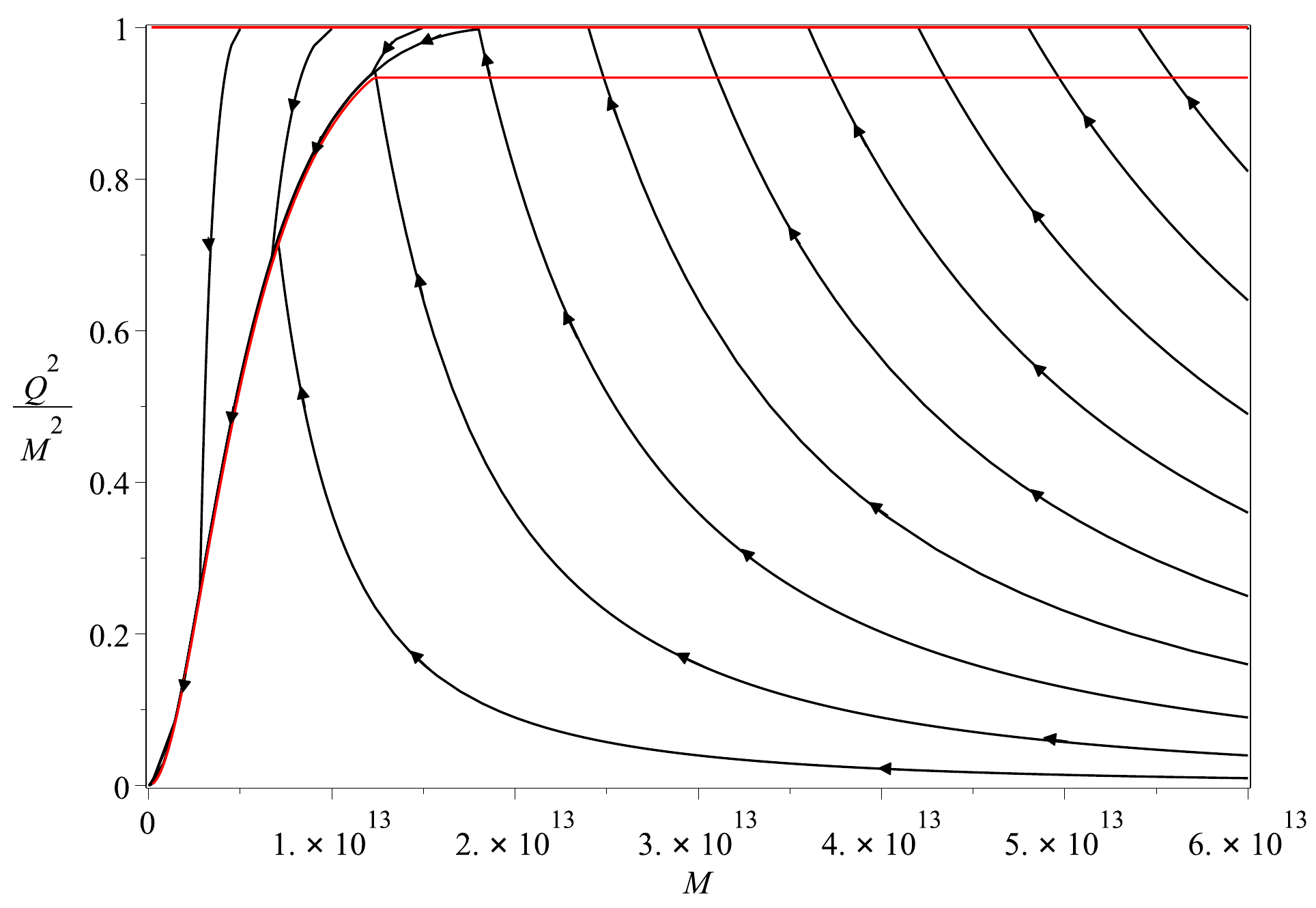}
\caption{The red (disconnected) curves satisfy $\frac{\partial}{\partial y}\left(\frac{\d T}{\d t}\right)^{-1} = 0$, where $y:=(Q/M)^2$. It approximates the attractor for $(Q/M)^2<3/4$, as well as for $(Q/M)^2 \approx 1$ at large $M$.
\label{attractor}}
\end{figure}

\begin{figure}[!h]
\centering
\includegraphics[width=3.4in]{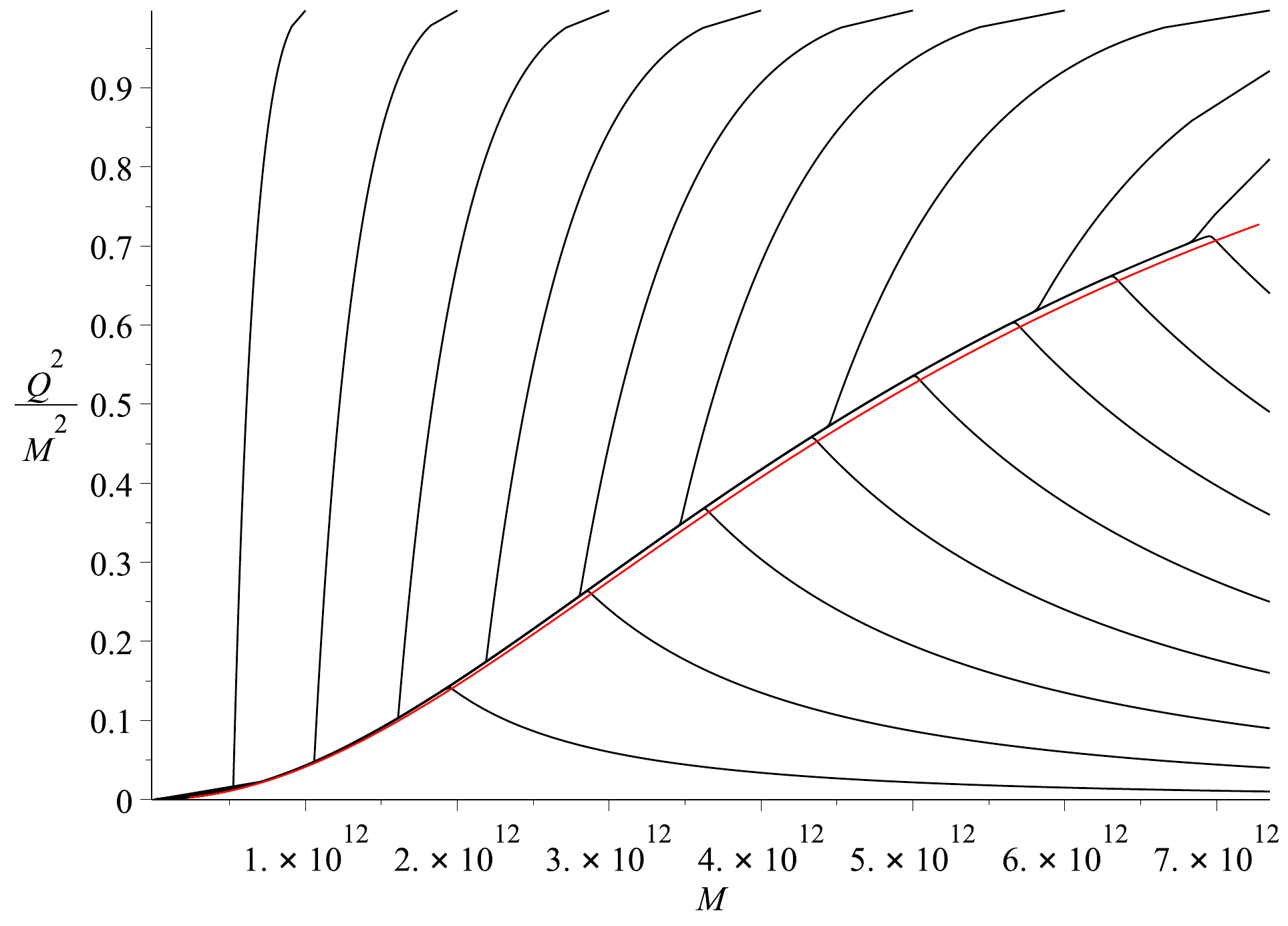}
\caption{The curve $\frac{\partial}{\partial y}\left(\frac{\d T}{\d t}\right) = 0$, or equivalently $\frac{\partial}{\partial y}\left(\frac{\d T}{\d t}\right)^{-1} \left(\frac{\d T}{\d t}\right)^{-1} = 0$, only approximates the attractor in the range $3/4<(Q/M)^2<1$, with better approximation at small value of $Q/M$. We have suppressed the arrows in this plot. 
\label{attractor-zoom}}
\end{figure}

Lastly, let us remark that the portion of the attractor curve  in the range $3/4<(Q/M)^2<1$ 
is well-approximated by the curve that satisfies
\begin{equation}
\frac{\partial}{\partial y}\left(\frac{\d T}{\d t}\right) = 0,
\end{equation}
or equivalently, 
\begin{equation}
\frac{\partial}{\partial y}\left(\frac{\d T}{\d t}\right)^{-1} = 0.
\end{equation}
The latter equation also gives the line $(Q/M)^2=1$, which coincides with the upper boundary of the positive specific heat region for large $M$. This is indicated in Fig.(\ref{attractor}). 
We shall refer to these curves as the ``lower curve'' and ``upper curve'', respectively.
Numerically the lower curve tends to a constant $(Q/M)^2 \approx 0.9336$, \emph{not} the lower boundary of the positive specific heat region $(Q/M)^2 =3/4$. The fact that this is only an approximation of the attractor can be seen from Fig.(\ref{attractor-zoom}), which is a zoomed-in version of part of the plot in Fig.(\ref{attractor}). The approximation gets better as $Q/M$ gets smaller. 

Note that the curves in the mass dissipation regime, whose charge-to-mass ratio is increasing, do not turn around when they hit the lower boundary of the positive heat region. Neither do they turn around when hitting the lower curve of  $\frac{\partial}{\partial y}\left(\frac{\d T}{\d t}\right)^{-1} = 0$. They continue to rise until hitting the upper boundary of the positive specific heat region, which happens to coincide with the upper curve of 
$\frac{\partial}{\partial y}\left(\frac{\d T}{\d t}\right)^{-1} = 0$.

\section{Conclusion}
The evolution of charged black holes can be rather subtle under Hawking evaporation, since the charge-to-mass ratio need not be monotonic in time. 
In this short letter, we showed that the attractor of an evaporating Reissner-Nordstr\"om black holes found by Hiscock and Weems \cite{HW} is characterized by the condition $\d M/\d t= \d Q/\d t$ and clarified some aspects of its nature. 
The attractor consists of two parts, which overlap:
\begin{itemize}
\item[(1)] $0 < (Q/M)^2 \lesssim 0.9336$: Well-approximated by the curve $\frac{\partial}{\partial y}\left(\frac{\d T}{\d t}\right)=0$, or equivalently,  $\frac{\partial}{\partial y}\left(\frac{\d T}{\d t}\right)^{-1}=0$.\\
\item[(2)] $3/4 <(Q/M)^2 <1$: Well-approximated by the boundary of the positive specific region. For $ (Q/M)^2 \approx 1$, it also satisfies $\frac{\partial}{\partial y}\left(\frac{\d T}{\d t}\right)^{-1}=0$.
\end{itemize}

It must be emphasized that positive specific heat region is not a necessary condition for the existence of an attractor. For example, evaporating dilaton charged black holes -- the (Gibbons-Maeda-)Garfinkle-Horowitz-Strominger solution \cite{gm, ghs} -- exhibit a similar attractor behavior \cite{1907.07490}. Nevertheless, their specific heat is always negative, since its Hawking temperature is identical to a Schwarzschild black hole of the same mass: $T=\hbar/(8 \pi M)$. Since the Hawking temperature is charge independent, the attractor curve for these GHS black holes cannot be approximated by $\frac{\partial}{\partial y}\left(\frac{\d T}{\d t}\right)=0$. In fact, the result of \cite{1907.07490} motivated the present work to seek a better understanding between the attractor of the Reissner-Nordstr\"om case and its specific heat. We found that these two do not actually coincide even for Reissner-Nordstr\"om black holes, although part of the attractor curve between $3/4 < y < 1$ does indeed come very close to the boundary of the positive specific heat region. An unsolved question remains: is there a deeper reason why these two come so close together yet do not coincide?

\begin{acknowledgments}
YCO thanks the National Natural Science Foundation of China (No.11705162) and the Natural Science Foundation of Jiangsu Province (No.BK20170479) for funding support. 
\end{acknowledgments}

\end{document}